\documentclass[11pt,a4paper]{article}
\usepackage{jcappub}
\usepackage{euscript}
\usepackage{amsfonts,latexsym}

\newcommand{\be}[1]{\begin{equation}\label{#1}}
\newcommand{\ee}{\end{equation}}
\newcommand{\ba}[1]{\begin{eqnarray}\label{#1}}
\newcommand{\ea}{\end{eqnarray}}
\newcommand{\rf}[1]{(\ref{#1})}
\newcommand{\nn}{\nonumber}

\begin{document}

\title{Remarks on mechanical approach to observable Universe}

\author{Maxim Eingorn$^{1}$} \author{and Alexander Zhuk$^{2}$}

\affiliation{$^{1}$CREST and NASA Research Centers, North Carolina Central University,\\ Fayetteville st. 1801, Durham, North Carolina 27707, U.S.A.\\}

\affiliation{$^{2}$Astronomical Observatory, Odessa National University,\\ Dvoryanskaya st. 2, Odessa 65082, Ukraine}

\emailAdd{maxim.eingorn@gmail.com} \emailAdd{ai.zhuk2@gmail.com}

\abstract{We consider the Universe deep inside the cell of uniformity. At these scales, the Universe is filled with inhomogeneously distributed discrete
structures (galaxies, groups and clusters of galaxies), which perturb the background Friedmann model. Here, the mechanical approach (Eingorn \& Zhuk, 2012) is
the most appropriate to describe the dynamics of the inhomogeneities which is defined, on the one hand, by gravitational potentials of inhomogeneities and, on
the other hand, by the cosmological expansion of the Universe. In this paper, we present additional arguments in favor of this approach. First, we estimate the
size of the cell of uniformity. With the help of the standard methods of statistical physics and for the galaxies of the type of the Milky Way and Andromeda,
we get that it is of the order of 190 Mpc which is rather close to observations. Then, we show that the nonrelativistic approximation (with respect to the
peculiar velocities) is valid for $z \lesssim 10$, i.e. approximately for 13 billion years from the present moment. We consider scalar perturbations and,
within the $\Lambda$CDM model, justify the main equations. Moreover, we demonstrate that radiation can be naturally incorporated into our scheme. This
emphasizes the viability of our approach. This approach gives a possibility to analyze different cosmological models and compare them with the observable
Universe. For example, we indicate some problematic aspects of the spatially flat models. Such models require a rather specific distribution of the
inhomogeneities to get a finite potential at any points outside gravitating masses. We also criticize the application of the Schwarzschild-de Sitter solution
to the description of the motion of test bodies on the cosmological background.}


\maketitle

\flushbottom

%%%%%%%%%%%%%%%%%%%%%%%%%%%%%%%%%%%%%%%%%%%%%%%%%%%%%%%%%%%%%%%%%

\section{\label{sec:1}Introduction}

\setcounter{equation}{0}

The standard hydrodynamical approach is a good tool for describing the growth of structure in the early Universe \cite{Mukhanov,Rubakov}. It works well in the
linear approximation. However, at the strongly nonlinear stage, when the primary protogalaxies were already formed and the Universe became highly inhomogeneous
on fairly small scales, it becomes inapplicable. It starts for $z$ less than a few dozens \cite{Gorbunov}. At this stage, to investigate the growth of
structure, the N-body simulation is commonly used (see, e.g., \cite{gadget2}). Such simulation is based on the Newtonian approach (in the evolving cosmological
background) described in \cite{Peebles}. Here, in the volume under consideration, matter is represented by a set of discrete point particles which correspond
to observable galaxies. On much bigger scales the Universe becomes on average homogeneous and isotropic with matter in the form of a perfect fluid. It is
important to define theoretically at which scales we should perform transition from the highly inhomogeneous mechanical distribution to the smooth
hydrodynamical one. Clearly, for different distributions we must apply different approaches. Next, it is also important to derive appropriate equations from
the first principles, i.e. from the General Relativity in our case.

The corresponding mechanical approach deep inside the cell of uniformity was proposed in our paper \cite{EZJCAP}. At these scales, the Universe is filled with
inhomogeneously distributed discrete structures (galaxies, groups and clusters of galaxies) which perturb the background Friedmann model. The mechanical
approach enabled us to obtain the gravitational potentials for an arbitrary number of randomly distributed inhomogeneities for models with conformally flat,
hyperbolic and spherical spaces within the $\Lambda$CDM model. In turn, with the expression for the gravitational potential, we can investigate the relative
motion of galaxies and the formation of the Hubble flows. In our paper \cite{EZK2}, we have applied this scheme to study the dynamics of our Local Group. Since
we get our equations from the first principles, we can generalize our analysis on the various alternative cosmological models and check their compatibility
with observations. This is an important advantage of our approach. At the present moment, there is a quite big number of alternative cosmological models which
try to describe the late Universe and to give an explanation of its late-time acceleration. They are, e.g., $\Lambda$CDM, non-linear $f(R)$ theories, Chaplygin
gas and quintessence models, Chevallier-Polarski-Linder model, etc. One of the main goals of our mechanical approach is to select viable models. In our
previous papers \cite{BUZ1,Laslo2,ENZ1}, we have demonstrated how this method works for some particular examples. To go further, we need to provide more
in-depth substantiation of the mechanical approach. For example, the notion of the cell of uniformity plays a significant role in this approach. So, we need to
provide its more concrete definition. Besides, peculiar velocities of inhomogeneities are nonrelativistic in our method. Therefore, we must determine up to
what time in the past we can use it. This is the subject of the present paper. Additionally, we demonstrate also how our approach can be used to investigate
the dynamics of galaxies and to study some problematic aspects of the nonrelativistic gravitational potential in the considered cosmological models.

The present paper is devoted to the further justification of the mechanical approach. In section 2, with the help of statistical physics, we  estimate the
scales of the cell of uniformity. We show that in the idealized Universe filled with the typical galaxies of the mass and size of the Milky Way or Andromeda,
the cell of uniformity size is of the order of 190 Mpc. This theoretical value is rather close to the observable data \cite{Geller_Huchra,Labini,Gorbunov}. In
the case of smaller typical galaxies (e.g., in one order of magnitude in size and two orders of magnitude in mass) we get 76 Mpc for the cell of uniformity
size, which is also rather large. In section 3, we show that the mechanical approach can be applied for $z \lesssim 10$ which corresponds approximately to 13
billion years from the present moment, i.e. almost the entire age of the Universe. In section 4, we justify our master equation \rf{4.11} for the
nonrelativistic gravitational potential within the $\Lambda$CDM model. Moreover, we demonstrate that radiation can be naturally included into our scheme. In
section 5, we, first, demonstrate that our approach can describe the growth of structure in the Universe. Then, we indicate some problematic aspects of the
flat models $\mathcal{K}=0$. We show that such models require a rather specific distribution of the inhomogeneities to get a finite potential at any points
outside gravitating masses. We also criticize the application of the Schwarzschild-de Sitter solution to the description of the motion of test bodies on the
cosmological background. One of the main drawbacks of such approach is that the Universe at large scales is the de Sitter one but not the Friedmann Universe.
Therefore, it does not take into account the matter in the Universe which contributes 31 \% into the total balance. This is the accuracy of this method. At the
very end of this section, we stress that the hyperbolic model $\mathcal{K}=-1$ is free of drawbacks inherent in the flat model $\mathcal{K}=0$. The main
results are summarized in concluding section 6.

%%%%%%%%%%%%%%%%%%%%%%%%%%%%%%%%%%%%%%%%%%%%%%%%%%%%%%%%%%%%%%%%%

\section{\label{sec:2}Cell of uniformity}

\setcounter{equation}{0}

According to the observations (see, e.g., \cite{Geller_Huchra,Labini,Gorbunov}), our present Universe is homogeneous at scales greater than 100-300 Mpc and its
space-time on these scales can be well described by the Friedmann-Robertson-Walker metrics with the matter content in the form of a perfect fluid. On smaller
scales the Universe is highly inhomogeneous and structured. We can see isolated galaxies, which can form clusters and superclusters. The observations also
strongly argue in favor of dark matter concentrated around these structures. Obviously, these visible and invisible isolated inhomogeneities can not be
represented in the form of a fluid. Therefore, hydrodynamics is not appropriate to describe their behavior on the considered scales. We need to create a
mechanical approach where dynamical behavior is defined by gravitational potentials. It is important to predict theoretically on what scales we can pass from
the mechanical approach into the hydrodynamical one. We will show now that such scales can be roughly estimated with the help of statistical physics. To show
it, we will follow the well known textbook by Klimontovich \cite{Klimontovich}.

Let us consider an idealized picture where the Universe is filled with galaxies which have masses
%of the order of
%masses
of the Milky Way and Andromeda: $m\sim 10^{12}M_{\odot}$ (where $M_{\odot}\approx 2\times10^{33}\, g$ is the mass of the Sun) and characteristic sizes $R_0
\sim 50$ kpc. For simplicity, we consider the galaxies as perfectly elastic balls with a diameter $R_0$. Taking into account the average density of the rest
mass in the Universe\footnote{\label{1}We make a snapshot of the idealized Universe at the present moment. Obviously, $\bar\rho_{\mathrm{ph}}$ depends on time.
This allows taking into account the dynamics of the Universe.} $\bar\rho_{\mathrm{ph}}\approx2.7\times10^{-30}\,g\cdot cm^{-3}$, then, following
\cite{Klimontovich} (see $\S$ 6 in chapter 7), we obtain at the present time the average intergalactic distance
$R_{\mathrm{mean}}=\left(m/\bar\rho_{\mathrm{ph}}\right)^{1/3}\sim 3$ Mpc and the mean free path $l_{\mathrm{mfp}}\sim
m/\left(\bar\rho_{\mathrm{ph}}R_0^2\right)\sim10^{4}$ Mpc. Here, we arrived at the first small parameter $\epsilon_1=
(R_0/R_{\mathrm{mean}})^3=\bar\rho_{\mathrm{ph}}R_0^3/m\sim5\times10^{-6}$ which is called the density parameter. Gases for which the density parameter is
small ($\epsilon_1 \ll 1$) are called rarefied ones. So, our idealized Universe is filled by a rarefied gas of galaxies with a good degree of accuracy. We can
consider such gas as a macroscopic system at scales $L\gg R_{\mathrm{mean}}$, that is at $L\gg 3$ Mpc for our case.

On what scale should we realize the averaging (smoothing) of the system to pass from the exact dynamic description to the averaged smoothed kinetic one? These
scales are usually called physically infinitely small and designated by $l_{\mathrm{ph}}$ and  $V_{\mathrm{ph}}=l_{\mathrm{ph}}^3$. They should satisfy the
following relations:
%%%%%%
\be{2.1}
R_0\ll R_{\mathrm{mean}}< l_{\mathrm{ph}}\ll l_{\mathrm{mfp}},\quad R_0^3\ll R_{\mathrm{mean}}^3< V_{\mathrm{ph}}\ll l_{\mathrm{mfp}}^3\, .
\ee
%%%%%%
According to \cite{Klimontovich}, $l_{\mathrm{ph}}\sim \sqrt{\epsilon_1}l_{\mathrm{mfp}}$ and $N_{\mathrm{ph}}=
\bar\rho_{\mathrm{ph}}V_{\mathrm{ph}}/m\sim1/\sqrt{\epsilon_1}\gg1$\footnote{\label{2}We can rewrite these formulas as follows: $l_{\mathrm{ph}}\approx
(R_{\mathrm{mean}}/R_0)^{1/2}R_{\mathrm{mean}}$ and $N_{\mathrm{ph}}\approx (R_{\mathrm{mean}}/R_0)^{3/2}$. Therefore, both of these values increase with
decreasing of $\epsilon_1$.}. The latter value is the number of particles/galaxies in $V_{\mathrm{ph}}$. Obviously, $N_{\mathrm{ph}}$ should be much bigger
than 1. In the opposite case the averaged (over $V_{\mathrm{ph}}$) local functions of the dynamical variables will not be considerably more smoothed than the
exact dynamic functions. In our case $l_{\mathrm{ph}}\sim22$ Mpc and $N_{\mathrm{ph}}\sim450$.

Next, we want to pass from the kinetic description to the hydrodynamical one. It takes place if we have the second small parameter $\epsilon_2 \equiv
l_{\mathrm{mfp}}/\tilde L \ll 1$ where $\tilde L$ is the representative physical length scale (in our case this is the physical size of the area of modeling
or, in other words, of the system of particles/inhomogeneities). This small parameter $\epsilon_2$ is called the hydrodynamic parameter or the Knudsen number.
Therefore, for our Universe $\tilde L\gg 10^4$ Mpc. We should stress that, from the statistical physics point of view, the hydrodynamical approach is available
only if we can  introduce the modeling area much greater than the mean free path of the particles. For the hydrodynamical physically infinitely small scale
(which is often referred to as the scale of a liquid particle) and the corresponding number of particles/galaxies inside it we have \cite{Klimontovich}:
%%%%%%
\be{2.2} l_{\mathrm{ph}}^{(\mathrm{hydro})}\sim \frac{\tilde L}{N^{1/5}},\quad
N_{\mathrm{ph}}^{(\mathrm{hydro})}\sim\frac{\bar\rho_{\mathrm{ph}}\left(l_{\mathrm{ph}}^{(\mathrm{hydro})}\right)^3}{m}= N^{2/5},\quad
N=\frac{\bar\rho_{\mathrm{ph}}\tilde L^3}{m}\, . \ee
%%%%%%%%
Let $\tilde L=10\cdot l_{\mathrm{mfp}}$, then for $l_{\mathrm{mfp}}\approx 10^4$ Mpc we get $l_{\mathrm{ph}}^{(\mathrm{hydro})}\approx 190$ Mpc and
$N_{\mathrm{ph}}^{(\mathrm{hydro})}\approx 277000$. Clearly, $l_{\mathrm{ph}}^{(\mathrm{hydro})}$ depends on $\tilde L$: it grows slowly as $\tilde L^{2/5}$.
Obviously, the larger $\tilde L$ we consider, the larger liquid particle we get and the more precise the hydrodynamical approach is. For our idealized model
the smallest possible cosmological liquid particle has the size of the order of 190 Mpc. This is the smallest scale of averaging to pass to the hydrodynamical
consideration. In other words, this is the cell of uniformity size. We see that our theoretical estimate, in spite of a rather simple model, is close to the
mentioned above experimental values \cite{Geller_Huchra,Labini,Gorbunov}. Hence, at the present time, the hydrodynamical approach is valid at scales greater
than $l_{\mathrm{ph}}^{(\mathrm{hydro})}$. At the smaller scales we should apply the mechanical approach (or at least the kinetic description at scales larger
than approximately 22 Mpc).

To conclude this section, it is of interest to see how the above results depend on the choice of the parameters of the statistical system, e.g. on the mass $m$
of particles/galaxies and their size $R_0$. Let these parameters be $m\sim 10^{10}M_{\odot}$ and $R_0\sim 5$ kpc (these are typical upper parameters for the
dwarf galaxies). Then, we get the estimates $R_{\mathrm{mean}}\sim0.6$ Mpc, $l_{\mathrm{mfp}}\sim 10^4$ Mpc and $\epsilon_1\sim 5\times10^{-7}$. Therefore, in
this case the characteristic scale of the transition from the exact dynamical description to the kinetic one is $l_{\mathrm{ph}}\sim7$ Mpc with the
corresponding number of particles $N_{\mathrm{ph}}\sim1420$ inside $V_{\mathrm{ph}}$. As we can see, the mean free path remains the same as before. Therefore,
we can choose the same Knudsen number $\epsilon_2$ and obtain for the hydrodynamical liquid particle $l_{\mathrm{ph}}^{(\mathrm{hydro})}\sim 76$ Mpc and
$N_{\mathrm{ph}}^{(\mathrm{hydro})}\sim 1745000$. Obviously, these values are smaller than in the previous example of the greater galaxies but they are still
rather big. It is worth noting that for both of these examples, the size of the area of modeling $\tilde L$ should be much greater than $10^4$ Mpc which
roughly corresponds to the size of the presently observable Universe. It means that the considered systems can be treated as the hydrodynamical ones if we
model them in the areas which exceed the observable Universe, that may make this procedure problematic. However, there is no any problem to apply the kinetic
approach at scales greater than 22 Mpc and 7 Mpc for the first and second examples, respectively. At such scales, we can already perform some averaging and
treat our systems as a fluid (not to be confused with the hydrodynamical fluid!) and calculate, e.g., the corresponding correlation functions.

\section{\label{sec:3}$\Lambda$CDM background model and mechanical approach}

\setcounter{equation}{0}

As we have seen above, on scales greater than $l_{\mathrm{ph}}^{(\mathrm{hydro})}$ the Universe may be considered homogeneous and isotropic and described by
the Friedmann-Robertson-Walker (FRW) metrics
%%%%%
\be{3.1}
ds^2=a^2\left(d\eta^2-\gamma_{\alpha\beta}dx^{\alpha}dx^{\beta}\right)=a^2\left(d\eta^2-\frac{\delta_{\alpha\beta}dx^\alpha
dx^\beta}{\left[1+\cfrac{1}{4}\mathcal K\left(x^2+y^2+z^2\right)\right]^2}\right)\, ,
\ee
%%%%%%%
where $\mathcal K=-1,0,+1$ for open, flat and closed Universes, respectively. The Friedmann equations for this metrics in the case of the $\Lambda$CDM model read
%%%%%%%
\be{3.2}
\frac{3\left({\mathcal H}^2+\mathcal K\right)}{a^2}=\kappa\overline{T}_{0}^0+\Lambda
\ee
%%%%%%%
and
%%%%%%%
\be{3.3}
\frac{2{\mathcal H}'+{\mathcal H}^2+\mathcal K}{a^2}=\Lambda\, ,
\ee
%%%%%%
where ${\mathcal H}\equiv a'/a\equiv (da/d\eta)/a$ and $\kappa\equiv 8\pi G_N/c^4$ ($c$ is the speed of light and $G_N$ is the Newton's gravitational
constant). Hereafter, the Latin indices $i,k,=0,1,2,3$ and the Greek indices $\alpha,\beta=1,2,3$. $\overline T^{ik}$ is the energy-momentum tensor averaged
over the volume $V_{\mathrm{ph}}^{(\mathrm{hydro})}$. For the late stages of the Universe evolution, we neglect the contribution of radiation if not stated
otherwise. For the pressureless dustlike matter, the energy density $\overline T^{0}_{0} =\overline \rho c^2/a^3$ is the only nonzero component (below, we will
show with what accuracy this statement is true). $\overline \rho$ is a constant which we define below. It is worth noting that in the case $\mathcal K =0$ the
comoving coordinates $x^{\alpha}$ may have a dimension of length, but then the conformal factor $a$ is dimensionless, and vice versa. However, in the cases
$\mathcal K=\pm 1$ the dimension of $a$ is fixed. Here, $a$ has a dimension of length and $x^{\alpha}$ are dimensionless. For consistency, we will follow this
definition for $\mathcal K=0$ as well.

Conformal time $\eta$ and synchronous time $t$ are connected as $cdt=a d\eta$. Therefore, eqs. \rf{3.2} and \rf{3.3}, respectively, take the form
%%%%%%
\be{3.4}
H^2=\left(\frac{\dot a}{a}\right)^2=\frac{\kappa\overline\rho c^4}{3a^3}+\frac{\Lambda c^2}{3}-\frac{\mathcal K
c^2}{a^2}=H_0^2\left(\Omega_{M}\frac{a_0^3}{a^3}+\Omega_{\Lambda}+\Omega_{\mathcal K}\frac{a_0^2}{a^2}\right)\, ,
\ee
%%%%%
and
%%%%%
\be{3.5}
\frac{\ddot a}{a}=-\frac{\kappa\overline\rho c^4}{6a^3}+\frac{\Lambda c^2}{3}=H_0^2\left(-\frac{1}{2}\Omega_{M}\frac{a_0^3}{a^3}+\Omega_{\Lambda}\right)\, ,
\ee
%%%%%
where $a_0$ and $H_0$ are the values of the conformal factor $a$ and the Hubble parameter $H\equiv \dot a/a\equiv (da/dt)/a$ at the present time $t=t_0$
(without loss of generality, we can put $t_0=0$), and we introduced the standard density parameters:
%%%%%%%
\be{3.6} \Omega_M=\frac{\kappa\overline\rho c^4}{3H_0^2a_0^3},\quad \Omega_{\Lambda}=\frac{\Lambda c^2}{3H_0^2},\quad\Omega_{\mathcal K}=-\frac{\mathcal K
c^2}{a_0^2H_0^2}\, ,\ee
%%%%%%%
therefore
%%%%%%
\be{3.7}
\Omega_M+\Omega_{\Lambda}+\Omega_{\mathcal K}=1\, .
\ee
%%%%%%%%

The first protogalaxies were formed for $z$ less than a few dozens \cite{Gorbunov}. As we will show below, our nonrelativistic approach works well for
red-shifts $z\lesssim10$. At this stage, the formation of inhomogeneities (galaxies and clusters of galaxies) has been generally
completed\footnote{\label{3}This is true up to collision/merging of the formed galaxies.}. The energy-momentum tensor components for such inhomogeneities read
\cite{Landau}:
%%%%%%
\be{3.8}
T^{ik}=\sum_p\frac{m_p c^2}{(-g)^{1/2}[\eta]}\frac{dx^i}{ds}\frac{dx^k}{ds}\frac{ds}{d\eta}\delta({\bf r}-{\bf r}_p)=
\sum\limits_p\frac{m_p c^2}{(-g)^{1/2}[\eta]}\frac{dx^i}{d\eta}\frac{dx^k}{d\eta}\frac{d\eta}{ds}\delta({\bf r}-{\bf r}_p)\, ,
\ee
%%%%%%
where $m_p$ is the mass of p-th inhomogeneity and $[\eta]$ indicates that the determinant is calculated from the metric coefficients defined with respect to
conformal time $\eta$. Taking into account peculiar velocities of inhomogeneities, we get
%%%%%%
\ba{3.9} T^0_0&=&\frac{\sqrt{\gamma}\rho c^2}{\sqrt{-g}}\frac{g_{00}+\tilde v^{\gamma}g_{0\gamma}}{\sqrt{g_{00}+2g_{0\alpha}\tilde
v^{\alpha}+g_{\alpha\beta}\tilde v^{\alpha}\tilde
v^{\beta}}}\, ,\\
\label{3.10}T_{\alpha}^0&=&\frac{\sqrt{\gamma}\rho c^2}{\sqrt{-g}}\frac{g_{0\alpha}+\tilde v^{\beta}g_{\alpha\beta}}{\sqrt{g_{00}+2g_{0\mu}\tilde v^{\mu}+
g_{\mu\nu}\tilde v^{\mu}\tilde
v^{\nu}}}\, ,\\
\label{3.11}T_{\beta}^{\alpha}&=&\frac{\sqrt{\gamma}\rho c^2}{\sqrt{-g}}\frac{\tilde v^{\alpha}\left(g_{0\beta}+\tilde
v^{\gamma}g_{\gamma\beta}\right)}{\sqrt{g_{00}+ 2g_{0\mu}\tilde v^{\mu}+g_{\mu\nu}\tilde v^{\mu}\tilde v^{\nu}}}\, , \ea
%%%%%%%
where $\gamma$ is the determinant of the metrics $\gamma_{\alpha\beta}$ and we introduce the rest mass density
%%%%%%
\be{3.12}
\rho=\frac{1}{\sqrt{\gamma}}\sum_p m_p \delta({\bf r}-{\bf r}_p)
\ee
%%%%%%
and $\tilde v^{\alpha}=dx^{\alpha}/d\eta=(a/c)dx^{\alpha}/dt\equiv av^{\alpha}/c\equiv v^{\alpha}_{\mathrm{ph}}/c$. Here, $v^{\alpha}_{\mathrm{ph}}$ is the
physical peculiar velocity. For the background metrics \rf{3.1}, these expressions take the form, respectively:
%%%%%%
\be{3.13}
T_0^0=\frac{\rho c^2}{a^3}\frac{1}{\sqrt{1-\gamma_{\alpha\beta}\tilde v^{\alpha}\tilde v^{\beta}}},\quad T_{\alpha}^0=-\frac{\rho
c^2}{a^3}\frac{\tilde v^{\beta}\gamma_{\alpha\beta}}{\sqrt{1-\gamma_{\mu\nu}\tilde v^{\mu}\tilde v^{\nu}}},\quad T_{\beta}^{\alpha}=-\frac{\rho
c^2}{a^3}\frac{\tilde v^{\alpha}\tilde v^{\gamma}\gamma_{\gamma\beta}}{\sqrt{1-\gamma_{\mu\nu}\tilde v^{\mu}\tilde v^{\nu}}}\, .
\ee
%%%%%%
Let us consider now a simplified model where all particles have the same in magnitude peculiar velocity $\tilde v=|{\bf \tilde v}|=\sqrt{\gamma_{\mu\nu}\tilde
v^{\mu}\tilde v^{\nu}}$. Then, averaging formulas \rf{3.13} over the volume  $V_{\mathrm{ph}}^{(\mathrm{hydro})}$, we obtain
%%%%%%%
\be{3.14} \overline{T_0^0}=\frac{\overline\rho c^2}{a^3}\frac{1}{\sqrt{1-\tilde v^2}}\equiv \overline\varepsilon_{\mathrm{ph}},\quad
\overline{T_{\alpha}^0}=0,\quad \overline{T_{\beta}^{\alpha}}=-\frac{\overline\rho c^2}{3a^3}\frac{\tilde v^2\delta_{\beta}^{\alpha}}{\sqrt{1-\tilde v^2}}
\equiv - \overline p_{\mathrm{ph}}\delta_{\beta}^{\alpha}\, , \ee
%%%%%%%
where $\overline \rho$ is the average rest mass density. These equations demonstrate that in the case of nonrelativistic peculiar velocities $\tilde v \ll 1$,
we have $\overline\varepsilon_{\mathrm{ph}} \approx \overline\rho c^2 /a^3$, $\overline p_{\mathrm{ph}} \approx 0$ and the energy density $\overline T^{0}_{0}
=\overline \rho c^2/a^3$ is the only nonzero component of the energy-momentum tensor. At the present time, the typical peculiar velocity is $v_{\mathrm{ph}}
\sim 300$ km/s, i.e. $\tilde v \sim 10^{-3} \ll 1$.

From the conservation equation $d(\overline\varepsilon_{\mathrm{ph}} a^3) + \overline p_{\mathrm{ph}}d(a^3)=0$ (where $\overline\varepsilon_{\mathrm{ph}}$ and
$\overline p_{\mathrm{ph}}$ are defined in \rf{3.14}) we can easily get
%%%%%%%
\be{3.15}
\tilde v^2=\frac1{1+C^2a^2}\, ,
\ee
%%%%%%%
where $C>0$ is a constant of integration. It is worth noting that an alternative derivation of this formula is presented in the book \cite{Gorbunov} (see
chapter 2.4). Hereinafter, we consider the mechanical approach where inhomogeneities have nonrelativistic velocities. Let us estimate now up to what times in
past such approach is valid, in other words, up to what red-shifts $z$ the peculiar velocities $\tilde v\lesssim 10^{-2}$. It can be easily found from the
above equation that it takes place for $z\lesssim 10$, i.e. approximately the last 13 billion years! Therefore, for $z\lesssim10$ we can neglect peculiar
velocities in \rf{3.14} with very good accuracy. Of course, our mechanical approach will remain valid in future. Strictly speaking, for earlier times with
$z\gg 10$ peculiar velocities should be taken into account.

\section{\label{sec:4}Perturbations}

\setcounter{equation}{0}

Obviously, the considered above inhomogeneities in the Universe result in scalar perturbations of the metrics \rf{3.1}. In the conformal Newtonian gauge, such
perturbed metrics is \cite{Mukhanov,Rubakov}
%%%%%%%
\be{4.1}
ds^2\approx a^2\left[(1+2\Phi)d\eta^2-(1-2\Psi)\gamma_{\alpha\beta}dx^{\alpha}dx^{\beta}\right]\, ,
\ee
%%%%%%%
where scalar perturbations $\Phi$ and $\Psi$ depend on all space-time coordinates $\eta,x,y,z$ and satisfy equations
%%%%%
\be{4.2}
\triangle\Psi-3{\mathcal H}(\Psi'+{\mathcal H}\Phi)+3\mathcal K \Psi=\frac{1}{2}\kappa a^2\delta T_{0}^0\, ,
\ee
%%%%%%%
%%%%%%%
\be{4.3}
\frac{\partial}{\partial x^{\beta}}(\Psi'+{\mathcal H}\Phi)=\frac{1}{2}\kappa a^2\delta T_{\beta}^0\, ,
\ee
%%%%%%%
%%%%%%%
\ba{4.4} \left[\Psi''+{\mathcal H}(2\Psi+\Phi)'+\left(2{\mathcal H}'+{\mathcal H}^2\right)\Phi+\frac{1}{2}\triangle(\Phi-\Psi)-\mathcal
K\Psi\right]\delta^{\alpha}_{\beta}\nn\\ -\frac{1}{2}\gamma^{\alpha\sigma}\left(\Phi-\Psi\right)_{;\sigma;\beta}=-\frac{1}{2}\kappa a^2\delta
T_{\beta}^{\alpha}\, , \ea
%%%
where, according to Eqs. \rf{3.9}-\rf{3.11},
%%%%%%%
\be{4.5}
\delta T^0_0=\delta\left(\frac{\sqrt{\gamma}\rho c^2\sqrt{g_{00}}}{\sqrt{-g}}\right),\quad \delta T_{\alpha}^0=\frac{\sqrt{\gamma}\rho c^2}{\sqrt{-g}}\frac{\tilde
v^{\beta}g_{\alpha\beta}}{\sqrt{g_{00}}},\quad \delta T_{\beta}^{\alpha}=\frac{\sqrt{\gamma}\rho c^2}{\sqrt{-g}}\frac{\tilde v^{\alpha}\tilde
v^{\gamma}g_{\gamma\beta}}{\sqrt{g_{00}}}\, .
\ee
%%%%%%%
The Laplace operator $\triangle$ and the covariant derivatives are defined with respect to the metrics $\gamma_{\alpha\beta}$. Obviously, in the second and
third expressions in Eq. \rf{4.5}, the metrics $g_{ik}$ is the background one. Following the standard argumentation (see, e.g., \cite{Mukhanov,Rubakov}), we
can put $\Phi=\Psi$. In Eq. \rf{4.5} the peculiar velocities $\tilde v^{\alpha}=dx^{\alpha}/d\eta$ are of the first order of smallness relative to their zero
background values. On the other hand, $\Phi$ is the first order perturbation of metrics and is connected with the nonrelativistic gravitational potentials of
the inhomogeneities.

The following important point should be noted. We consider the weak field limit. It means that gravitational fields of the inhomogeneities are weak ($\Phi\ll
1$). Besides, their peculiar velocities are nonrelativistic ($\tilde v \ll 1 \Leftrightarrow v_{\mathrm{ph}} \ll c$).  Obviously, weak gravitational fields
represent a requirement for nonrelativistic motions of test bodies, but the opposite is not always true. Very light relativistic bodies/particles may produce a
weak gravitational field. Therefore, we can consider these two conditions independently of each other. Concerning our mechanical approach, it means that we can
consider Eqs. \rf{4.2}-\rf{4.4} in the first order approximation with respect to $\Phi$ and in the zero order approximation with respect to $\tilde v$. Such
approach is fully consistent with \S 106 in \cite{Landau} where the gravitational field of an arbitrary number of massive discrete bodies was found in the weak
field approximation. It was shown that the nonrelativistic gravitational potential is defined by the positions of the inhomogeneities but not by their
velocities (see Eq. (106.11) in this book). In the case of an arbitrary number of dimensions, a similar result was obtained in \cite{EZ3}. Thus, to describe
the dynamical evolution of the system, we divide the problem into two phases. First, we obtain the gravitational potential of the considered system. Then, this
potential is used to determine the dynamical behavior of the system (e.g., for the numerical simulation).

Such scenario was realized in our papers \cite{EZJCAP,EZK2}. Now, we want to add some comments. First, it can be easily seen that the second and third formulas
in Eq. \rf{4.5} are of the first and second order of smallness with respect to the peculiar velocities. Hence, we can drop them in the right hand sides of Eqs.
\rf{4.3} and \rf{4.4}, respectively:
%%%%%%%%
\ba{4.3a}
\frac{\partial}{\partial x^{\beta}}(\Phi'+{\mathcal H}\Phi)&=&0\, ,\\
\label{4.4a}
\Phi''+3{\mathcal H}\Phi'+\left(2{\mathcal H}'+{\mathcal H}^2\right)\Phi-\mathcal K \Phi&=&0\, .
\ea
%%%%%%%%
This enables us to get a solution of \rf{4.3a} in the form
%%%%%
\be{4.6}
\Phi(\eta,{\bf r})=\frac{\varphi({\bf r})}{c^2a(\eta)}\, ,
\ee
%%%%%
where $\varphi({\bf r})$ is a function of all spatial coordinates and we have introduced $c^2$ in the denominator for convenience. This function behaves as
$\varphi({\bf r})\sim 1/r$ in the vicinity of an inhomogeneity \cite{EZJCAP}, and the nonrelativistic gravitational potential $\Phi(\eta,{\bf r})\sim 1/(a
r)=1/R$, where $R=ar$ is the physical distance. Hence, $\Phi$ has the correct Newtonian limit near the inhomogeneities. For $\Phi$ in the form of \rf{4.6}, Eq.
\rf{4.2} takes the form
%%%%%
\be{4.7}
\triangle\Phi+3\mathcal K \Phi=\frac{1}{2}\kappa a^2\delta T_{0}^0\, .
\ee
%%%%%%%
Taking into account the perturbed metrics \rf{4.1}, we get in the first approximation
%%%%%
\be{4.8}
\delta T_{0}^0=\frac{\delta\rho c^2}{a^3}+\frac{3\overline{\rho}c^2\Phi}{a^3}=\frac{\delta\rho c^2}{a^3}+\frac{3\overline{\rho}\varphi}{a^4}\, ,
\ee
%%%%%%%
where $\delta\rho$ is the difference between real and average comoving rest mass densities:
%%%%%
\be{4.9}
\delta\rho = \rho-\overline\rho\, .
\ee
%%%%%
In the right hand side of Eq. \rf{4.8}, the second term is proportional to $1/a^4$ and should be dropped because we consider the $\Lambda$CDM model with
nonrelativistic matter. This is the accuracy of our approach. Hence, the perturbation of the energy-density reads
%%%%%
\be{4.10}
\delta T_{0}^0=\frac{\delta\rho c^2}{a^3}\, .
\ee
%%%%%%
Therefore, Eq. \rf{4.7} with respect to the function $\varphi$ takes the form
%%%%%%
\be{4.11}
\triangle\varphi+3\mathcal K\varphi=4\pi G_N (\rho-\overline\rho)\, .
\ee
%%%%%%
In the flat case $\mathcal K=0$, this equation  coincides (up to evident redefinition) with the equation (7.9) in \cite{Peebles} and the corresponding equation
in section 2.1 in \cite{gadget2}. Moreover, Eqs. (7.9) and (7.14) in \cite{Peebles} also demonstrate that $\Phi \sim 1/a$.

Eq. \rf{4.11} was solved in the $\Lambda$CDM model for $\mathcal K =0,\pm 1$ in our paper \cite{EZJCAP}. The recent observations indicate in favor of a flat
model $\mathcal K=0$. However, small spatial curvatures are also not excluded. Therefore, we have considered all possible values of $\mathcal K$. We have shown
that the hyperbolic case $(\mathcal K=-1)$ has some advantage with respect to both flat $(\mathcal K =0)$ and spherical $(\mathcal K=+1)$ cases. We
demonstrated that in the flat case
%, to avoid the famous Neumann-Seeliger paradox (see, e.g., the review \cite{Norton}),
the distribution of the gravitating bodies should be rather specific.
%Moreover, average gravitational potential of all masses in the Universe is not equal to zero, although physically reasonable to assume its %zero value.
If we try to distribute masses arbitrarily, we can arrive at a contradiction in the form of divergent gravitational potentials in points where masses are
absent. For example, in the case of the periodic distribution of the gravitating masses $m$ with periods $l_1,l_2$ and $l_3$ along x-,y- and z-axis,
respectively, the solution of \rf{4.11} (where $\mathcal K=0$) is \cite{periodic}:
%%%%%
\be{4.11a}
\varphi=-\frac{G_Nm}{\pi l_1 l_2 l_3}\sum_{k_1=-\infty}^{+\infty} \sum_{k_2=-\infty}^{+\infty}\sum_{k_3=-\infty}^{+\infty} \frac{\cos\left(2\pi
k_1x/l_1\right)\cos\left(2\pi k_2y/l_2\right)\cos\left(2\pi k_3z/l_3\right)}{k_1^2/l_1^2+k_2^2/l_2^2+ k_3^2/l_3^2}\, ,
\ee
%%%%%
where $k_1^2+k_2^2+k_3^2\neq0$. Here, at the point $x=y=z=0$, i.e. where the gravitating mass is, we get the usual Newtonian divergence: $\varphi \sim 1/r$.
However, this potential is also divergent if we fix at a nonzero value any of the coordinates, e.g., $x\neq0$, but allow simultaneously $y,\,z \to 0$. This
means that this potential is divergent at points where the gravitating masses are absent. Clearly, this is an unphysical result.
%For the periodic distribution of masses, we also arrive at the similar problem \cite{periodic}.
In the spherical space case, there is also unphysical divergence of the potential in antipodal points \cite{EZJCAP}. The hyperbolic case is
free from these drawbacks. However, it is worth mentioning here that adding an exotic matter can improve the situation. For example, in \cite{BUZ1} we have
shown that the presence of the frustrated network of cosmic strings results in reasonable gravitational potentials for any value of $\mathcal K$.

Formally, the function $\varphi({\bf r})$ does not depend on time. However, gravitating inhomogeneities are not fixed bodies. They move with respect to each
other. Therefore, this function depends on time implicitly. Let us demonstrate that even if we take into account such dependence on time, the omitted term in
Eq. \rf{4.2} will be negligible compared with the remaining ones. Keeping in mind that
%%%%%%
\be{4.12} 3\mathcal H(\Phi'+\mathcal H\Phi)=3\mathcal H\frac{1}{c^2a}\frac{\partial\varphi}{\partial
x^{\alpha}}\frac{dx^{\alpha}}{d\eta}=\frac{3aH}{c}\,\frac{\partial\Phi}{\partial x^{\alpha}}\,\frac{av^{\alpha}}{c}\, , \ee
%%%%%%
we obtain at the present time an estimate
%%%%%%
\be{4.13} \left|\frac{-3\mathcal H(\Phi'+\mathcal H\Phi)}{\triangle\Phi}\right|\sim \frac{(3a_0H_0/c)\,
(\Phi/r)}{\Phi/r^2}\frac{av}{c}=\frac{3R_{\mathrm{ph}}H_0}{c}\frac{v_{\mathrm{ph}}}{c}\sim 2\times 10^{-6}\, , \ee
%%%%%%
where $R_{\mathrm{ph}}\sim R_{\mathrm{mean}}\sim 3$ Mpc is the characteristic scale of variation of the function $\Phi$ (which is of the order of the average
intergalactic distance at the present time). For the peculiar velocities we take $v_{\mathrm{ph}}/c\sim 10^{-3}$, and the Hubble constant $H_0\approx70\,
\mbox{km/sec/Mpc}\approx 2.3\times 10^{-18}\mbox{sec}^{-1}$. Hence, the term $3\mathcal H(\Phi'+\mathcal H\Phi)$ is really much less than $\triangle \Phi $.
Additionally, Eq. \rf{4.12} demonstrates that this omitted term is of the first order of smallness with respect to velocities (times the other small parameter
$\Phi$)\footnote{\label{4}Moreover, taking into account that $\Phi \sim O(1/c^2)$, we get that the expression \rf{4.12} is of the order of $O(1/c^4)$, while
all the other terms in Eq. \rf{4.2} are of the order of $O(1/c^2)$.}.  Therefore, it should be dropped according to the considered accuracy.

With the help of the Friedmann equations, it can be easily verified that Eq. \rf{4.4a} is satisfied up to $1/a^4$ if $\Phi$ has the form \rf{4.6} (see
\cite{EZJCAP}). This corresponds to the accuracy of our approach because we consider nonrelativistic matter (see the text after Eq. \rf{4.9}).  Let us prove
now a more rigorous statement that Eq. \rf{4.4a} is satisfied up to $1/a^5$ if we include radiation into our model. In the presence of radiation, Eqs. \rf{4.2}
and \rf{4.4a} read
%%%%%%
\ba{4.14}
\triangle\Phi-3{\mathcal H}(\Phi'+{\mathcal H}\Phi)+3\mathcal K \Phi&=&\frac{1}{2}\kappa a^2\left(\delta T_{0}^0+\delta\varepsilon_{\mathrm{rad}}\right)\, , \\
\label{4.15}\Phi''+3{\mathcal H}\Phi'+\left(2{\mathcal H}'+{\mathcal H}^2\right)\Phi-\mathcal K \Phi&=&\frac{1}{2}\kappa a^2\delta p_{\mathrm{rad}} \, , \ea
%%%%%%
where $\delta\varepsilon_{\mathrm{rad}}$ and $\delta p_{\mathrm{rad}}=\delta\varepsilon_{\mathrm{rad}}/3$ are fluctuations of the energy density and pressure
of radiation, respectively. Eq. \rf{4.3a} preserves its form and results again in the nonrelativistic gravitational potential \rf{4.6}. The fluctuation of the
energy density of nonrelativistic matter $\delta T_{0}^0$ is given by Eq. \rf{4.8}. Therefore, from Eq. \rf{4.14} we obtain the relation
%%%%%%%%
\be{4.16}
\frac{3\overline\rho\varphi}{a^4}+\delta\varepsilon_{rad}=0\quad\Rightarrow\quad \delta\varepsilon_{rad}=-\frac{3\overline\rho\varphi}{a^4}
= -3\overline\rho_{\mathrm{ph}}c^2\Phi
\, .
\ee
%%%%%%%%
Taking into account the relations
%%%%%%
\be{4.17}
\Phi'=-\frac{\varphi a'}{c^2a^2}=-{\mathcal H}\Phi,\quad\Phi''=-{\mathcal H}'\Phi+{\mathcal H}^2\Phi\, ,
\ee
%%%%%%
we can rewrite Eq. \rf{4.15} in the following form:
%%%%%%%
\be{4.18}
\left({\mathcal H}'-{\mathcal H}^2-\mathcal
K\right)\Phi=\frac{1}{2}\kappa a^2\frac13\delta \varepsilon_{\mathrm{rad}}\, .
\ee
%%%%%%%
On the other hand, from the Friedmann equations we obtain
%%%%%%%
\be{4.19}
{\mathcal H}'-{\mathcal H}^2-\mathcal K = -\frac12 \kappa a^2 \left(\overline T^0_{0} + \frac43 \overline\varepsilon_{\mathrm{rad}} \right)\, ,
\ee
%%%%%%%
where $\overline T^0_0 = \overline \rho c^2/a^3$ and $\overline\varepsilon_{\mathrm{rad}}\sim 1/a^4$ are the energy densities of background nonrelativistic
matter and background radiation, respectively. Substituting \rf{4.16} and \rf{4.19} into \rf{4.18}, we arrive at the following relation
%%%%%%%
\be{4.20}
-\frac{\kappa a^2\overline{T}_{0}^0}{2}\Phi=\frac{1}{2}\kappa a^2 \frac{1}{3}\left(-\frac{3\overline\rho\varphi}{a^4}\right)\, ,
\ee
%%%%%%%
where in the left hand side we drop the background radiation term $\overline \varepsilon_{\mathrm{rad}}\Phi \sim 1/a^5$. It can be easily seen that this
equation is satisfied identically. It should be noted that, up to notation, Eq. \rf{4.16} coincides with the equation (40) in \cite{Vlasov} if we also neglect
the contribution of background radiation.

To conclude this section, let us demonstrate now how we can take into account the velocities of the inhomogeneities (in the first order of smallness) within
the scope of our approach. To do it, we consider the vector perturbations of the metrics:
%%%%%
\be{4.21}
ds^2\approx a^2\left[(1+2\Phi)d\eta^2+2S_{\alpha}d\eta dx^{\alpha}-(1-2\Phi)\gamma_{\alpha\beta}dx^{\alpha}dx^{\beta}\right]\, .
\ee
%%%%%
Let us consider, for a simplicity, the flat case $\mathcal K=0$. Then, taking into account Eq. \rf{4.5}, we get
%%%%%%
\be{4.22}
\triangle S_{\alpha}=\frac{16\pi G_N}{c^4}a^2\delta T_{\alpha}^0+\ldots=-\frac{16\pi
G_N}{c^2}\frac{\rho}{a}\frac{av^{\alpha}}{c}+\ldots\, ,
\ee
%%%%%%
where the dots denote terms of the same order of magnitude which follow from the implicit dependence of $\varphi$ on $t$. Therefore, for a
particle/inhomogeneity with the mass $m$ we get
%%%%%%
\be{4.23} S_{\alpha}=\left(\frac{4}{c^2}\frac{G_Nm}{ar}\frac{av^{\alpha}}{c}+\ldots\right)\sim \left(\Phi \frac{av^{\alpha}}{c}+\ldots\right) \sim
\left[O\left(\frac{1}{c^3}\right)+\ldots\right]\, . \ee
%%%%%%
It is worth noting that this formula is similar to the first expression on the right hand side of the equation for $h_{0\alpha}$ in \cite{Landau} (see the
equation just after (106.14)). As we can see, the inclusion into consideration of the peculiar velocities results in terms of the higher order of smallness for
the metric coefficients perturbations. These terms are proportional to the product of two small parameters: $\Phi$ and $\tilde v$, and this product is of the
order of $O(1/c^3)$. In the zero order approximation with respect to $\tilde v$, all metrics perturbations are of the order of $O(1/c^2)$. Therefore, peculiar
velocities lead to higher order terms which do not affect significantly the dynamics of inhomogeneities on the considered stage of the Universe evolution.
%%%%%%%%%%%%%%%%%%%%%%%%%%%%%%%%%%%%%%%%%%%%%%%%%%%%%%%%%%%%%%%%%%%%%%%%%%%%%%%%%%%%%%%%%%%%%%%%%%%%%

\section{\label{sec:5}Dynamics of inhomogeneities}

\setcounter{equation}{0}

Knowing the gravitational potential generated by gravitating masses, we can now determine the dynamical behavior of the inhomogeneities. Their motion is
defined by the Lagrange equations. To get the Lagrange function of a test mass $m$, let us remind that the action for this test body can be written in the
following form \cite{Landau}:
%%%%%%%
\be{5.1}
S=-mc\int ds\approx-mc\int\left\{\left(1+2\Phi\right)c^2-a^2\left(1-2\Phi\right)v^2\right\}^{1/2}dt\, ,\quad
v^2=\gamma_{\alpha\beta}\dot{x}^{\alpha}\dot{x}^{\beta}\, .
\ee
%%%%%%%
Hence, the corresponding Lagrange function has the form
%%%%%%
\be{5.2}
L\approx-mc^2\left\{1+2\Phi -a^2\frac{v^2}{c^2}(1-2\Phi)\right\}^{1/2}\approx-mc^2\left(1+\frac{\varphi}{ac^2}
-\frac{a^2v^2}{2c^2}\right)=-mc^2-\frac{m\varphi}{a}+
\frac{ma^2v^2}{2}\, ,
\ee
%%%%%%
where we dropped the term $O\left(1/c^2\right)$. For nonrelativistic test masses, we can also drop the term $mc^2$. It can be easily verified that the Hamilton
function for the system of inhomogeneities exactly coincides with the formula (1) in the cosmological simulation code GADGET-2 \cite{gadget2}. Below, for
simplicity, we consider the case of the flat Universe $(\mathcal K=0)$. Then, the Lagrange equation is
%%%%%%
\be{5.3}
\frac{d}{dt}\left(a^2{\bf v}\right)=-\frac{1}{a}\frac{\partial \varphi}{\partial {\bf r}}\, ,
\ee
%%%%%%%
where for the flat comoving space ${\bf v}=d{\bf r}/dt$. Clearly, on the scales less than the cell of uniformity size we can use Cartesian coordinates for any
signs of $\mathcal K$. Therefore, the simplification to the case $\mathcal K=0$ does not affect our following conclusions concerning the growth of structure.
In terms of the physical distance ${\bf R} =a{\bf r}$ and the physical velocity ${\bf V}\equiv d{\bf R}/dt=d(a{\bf r})/dt$, this equation reads
%%%%%%
\be{5.4}
\frac{d{\bf V}}{dt}=\frac{\ddot a}{a}{\bf R}-\frac{1}{a}\frac{\partial\varphi}{\partial{\bf R}}\, .
\ee
%%%%%%
We used this equation in \cite{EZK2} to describe the dynamics of our Local Group. Let us demonstrate now that this equation can be used also to determine the
growth of structure in the Universe. To show it, considering the physical velocity ${\bf V}$ as a function of time $t$ and physical spatial coordinates $\bf
R$, we rewrite Eq. \rf{5.4} as follows:
%%%%%%
\be{5.5} \frac{\partial {\bf V}}{\partial t}+\left({\bf V}\frac{\partial}{\partial {\bf R}}\right){\bf V}=\frac{\ddot a}{a}{\bf R}-\frac{\partial\tilde
\Phi}{\partial {\bf R}}\, , \ee
%%%%%%%
where the gravitational potential $\tilde \Phi\equiv \varphi/a$ satisfies the equation (see Eq. \rf{4.11} for $\mathcal K=0$)
%%%%%
\be{5.6}
\triangle_{{\bf R}}\tilde\Phi=4\pi G_N\frac{1}{a^3}(\rho-\overline\rho)=4\pi
G_N(\rho_{\mathrm{ph}}-\overline\rho_{\mathrm{ph}})\equiv 4\pi G_N\delta\rho_{\mathrm{ph}}\, ,
\ee
%%%%%
where the Laplace operator $\triangle_{\bf R} = \sum_{\alpha=1}^3\partial^2/(a\partial x^{\alpha})^2$ and the rest mass density $\rho$ is defined by \rf{3.12}
where $\gamma=1$. These equations must be supplemented with the continuity equation
%%%%%%
\be{5.7}
\frac{\partial \rho_{\mathrm{ph}}}{\partial t}+\frac{\partial}{\partial {\bf R}}\left(\rho_{\mathrm{ph}}{\bf V}\right)=0\, .
\ee
%%%%%%
Let us linearize Eqs. \rf{5.5} and \rf{5.7}. In the zero order approximation (with respect to linearization) $\tilde \Phi \equiv 0$ and
$\rho_{\mathrm{ph}}=\overline\rho_{\mathrm{ph}} (t)$. Then, from Eqs. \rf{5.5} and \rf{5.7} we obtain ${\bf V} =\overline {\bf V} = H {\bf R}$ (i.e the Hubble
law) and $\overline \rho_{\mathrm{ph}} \sim 1/a^3$. In the first order approximation with respect to the small perturbations $\delta\rho_{\mathrm{ph}}, \,
\tilde \Phi$ and $\delta {\bf V} = {\bf V}-\overline{\bf V}={\bf v}_{\mathrm{ph}}$, Eqs. \rf{5.5} and \rf{5.7} are reduced to the following ones:
%%%%%%
\ba{5.8}
\frac{\partial {\bf v}_{\mathrm{ph}}}{\partial t} + H{\bf v}_{\mathrm{ph}} + H \left({\bf R}\frac{\partial}{\partial{\bf R}}\right)
{\bf v}_{\mathrm{ph}} &=& - \frac{\partial\tilde \Phi}{\partial {\bf R}}\, ,\\
\label{5.9} \frac{\partial \delta}{\partial t} + H\left({\bf R}\frac{\partial}{\partial{\bf R}}\right)\delta + \frac{\partial}{\partial{\bf R}}{\bf
v}_{\mathrm{ph}}&=& 0\, , \ea
%%%%%%
where $\delta ({\bf R},t)\equiv \delta \rho_{\mathrm{ph}}({\bf R},t) / \overline \rho_{\mathrm{ph}}(t)$ is the density contrast and we use the Friedmann
equations. Taking into account that the speed of sound for dustlike matter is equal to zero, it can be easily verified that the system of equations \rf{5.6},
\rf{5.8} and \rf{5.9} exactly coincides with the equations (1.17a), (1.17c) and (1.17b) in \cite{Rubakov}, respectively, where these equations were used to
demonstrate the growth of the density contrast.

Now, we want to demonstrate some problematic aspects of the ${\mathcal K}=0$ model. In the previous section, we have already written about unphysical
divergencies in the case of the periodic distribution of the gravitating masses in the flat model. Formally, the solution of Eq. \rf{5.6} can be written in the
form (see, e.g., the formula (8.1) in \cite{Peebles}):
%%%%%%
\be{5.10}
\tilde \Phi ({\bf r}) = - G_N a^2 \int d {\bf r}'\frac{\rho_{\mathrm{ph}}({\bf r}')-\overline\rho_{\mathrm{ph}}}{|{\bf r}'-{\bf r}|}\, .
\ee
%%%%%%
To get a finite value of $\tilde \Phi ({\bf r})$ in the case of the infinite space (the infinite number of
particles), we need to assume some form of the spatial distribution of gravitating masses. One of such distributions was proposed in our paper  \cite{EZJCAP}.
We suppose that in the vicinity of each inhomogeneity the gravitational potential is defined by the mass of this inhomogeneity and is not affected by other
masses. We consider a simplified version where the inhomogeneities are approximated by point-like masses, which do not interact gravitationally with each
other. Further, we assume that each point-like mass $m_{0{i}}$ (here, we introduce the subscript $0$ to differ these gravitating masses from a test mass $m$)
is surrounded by an empty sphere of the radius $r_{0{i}}$ and this sphere, in turn, is embedded in a medium with the rest mass density $\overline\rho$. Such
supposition of the spatial distribution of matter provides the finiteness of the gravitational potential at any point of space and for an arbitrary number of
inhomogeneities. Then, for any gravitating mass $m_0$ the solution of \rf{5.6} is
%%%%%%
\ba{5.11}
\varphi&=&-\frac{G_Nm_0}{r}-\frac{G_Nm_0}{2r_0^3}r^2+\frac{3G_Nm_0}{2r_0}\, ,\quad r\le r_0\, ,\\
\label{5.12}\varphi&\equiv& 0\quad \Rightarrow\quad \frac{d\varphi}{dr}=0\, , \quad r\ge r_0\, ,
\ea
%%%%%%
where the matching condition gives
%%%%%
\be{5.13}
\quad r_0=\left(\frac{3m_0}{4\pi\overline{\rho}}\right)^{1/3}\, .
\ee
%%%%%%
The average value of the potential \rf{5.11} is $\overline{\varphi}=-3G_Nm_0/(10r_0)\neq0$. The formula \rf{4.16} connects the gravitational potential and the
fluctuations of the energy density of radiation. Obviously, the average value of the fluctuations of the energy density of radiation should be equal to zero.
Therefore, the average value of the gravitational potential must also be equal to zero. This is one of the main drawbacks of the expression \rf{5.11}.

It can be easily verified that, for the given mass distribution, a similar solution \rf{5.11} follows from the integral \rf{5.10}. Therefore, the Peebles
integral \rf{5.10} for some distributions can also result in nonzero average values of the gravitational potentials.

Eq. \rf{5.11} can be rewritten in the
form
%%%%%%%
\be{5.14}
\tilde \Phi = -\frac{G_Nm_0}{R} - \frac{2\pi G_N}{3}\overline\rho_{\mathrm{ph}} R^2 + 2\pi G_N \overline\rho_{\mathrm{ph}} R_0^2\, ,\quad R\le R_0\, ,
\ee
%%%%%%%
where $R_0=a r_0$. This equation demonstrates that the peculiar acceleration
%%%%%%%
\be{5.15} {\bf g} =-\frac{1}{a^2} \nabla_{\bf r}\varphi = -\nabla_{\bf R}\tilde \Phi = -\frac{G_Nm_0}{R^3}{\bf R} - \frac{4\pi
G_N}{3}\overline\rho_{\mathrm{ph}} {\bf R}\, ,\quad R\le R_0 \ee
%%%%%%%
is not reduced to a solely Newtonian expression in contrast to the formula (8.5) in \cite{Peebles}. Obviously, this is a consequence of the chosen distribution
of gravitating masses. The peculiar acceleration \rf{5.15} is equal to zero at $R_0$. Therefore, $R_0$ is the radius of zero acceleration (the radius of local
gravity). Outside this surface (and also outside other local gravity surfaces), test masses move according to the Hubble law: $d{\bf V}/dt = (\ddot{a}/a){\bf
R}$ (see Eq. \rf{5.4}).

We may also introduce another potential:
%%%%%%
\be{5.16} \Phi_{\mathrm{SdS}} = \tilde \Phi -\frac12 \frac{\ddot{a}}{a} R^2 = -\frac{G_N m_0}{R}-\frac{\Lambda c^2}{6}R^2 + 2\pi G_N
\overline\rho_{\mathrm{ph}} R_0^2\, ,\quad R\le R_0\, , \ee
%%%%%%
where in the latter equality we have used the Friedmann equation \rf{3.5} and the solution \rf{5.14}. Then, the Lagrange equation \rf{5.4} reads
%%%%%%
\be{5.17}
\frac{d{\bf V}}{dt}=-\frac{\partial}{\partial{\bf R}} \Phi_{\mathrm{SdS}}=-\frac{G_N m_0}{R^3}{\bf R}+\frac{\Lambda c^2}{3}{\bf R}\, ,\quad
R\le R_0\, .
\ee
%%%%%%
If we drop the condition $R\le R_0$, then this equation describes the motion of a test mass in the gravitational field associated with the Schwarzschild-de
Sitter solution. Sometimes, this equation is used to describe the motion of test bodies in the cosmological background \cite{CherninUFN2,BisChern,GibbEll}. We
would like to stress some weak points of this approach.

First of all, as we can see from \rf{5.17}, the dynamical behavior at large distances $R$ is mainly defined by the cosmological constant $\Lambda$:
%%%%%%
\be{5.17a}
\frac{d{\bf V}}{dt} \rightarrow \frac{\Lambda c^2}{3}{\bf R}\, .
\ee
%%%%%%
That is spacetime asymptotically corresponds to the de Sitter Universe. However, at large scales, our Universe is not the de Sitter one but the Friedmann
Universe where its dynamics is defined by both matter and dark energy. Therefore, at large distances from inhomogeneities test bodies should follow the Hubble
flows (up to peculiar velocities), and we should have asymptotically
%%%%%%
\be{5.18}
\frac{d{\bf V}}{dt} \rightarrow\frac{\ddot{a}}{a} {\bf R} = \left(-\frac{\kappa\overline\rho c^4}{6a^3}+\frac{\Lambda c^2}{3}\right){\bf R}\, .
\ee
%%%%%%
According to the recent observations (see, e.g., the Table 2 in \cite{Planck}), matter contributes approximately 31\ \% into the total balance. Hence, dropping
the contribution of matter corresponds to a decrease in accuracy of 31 percent. In our specific model considered above, the transition \rf{5.18} occurs in the
region outside the sphere of local gravity due to smooth cutoff of the potential $\tilde \Phi$ at $R=R_0$.

Taking into account the Friedmann equation \rf{3.5} and Eq. \rf{5.6}, it can be easily seen that in the case $\mathcal{K}=0$ the potential $\Phi_{\mathrm{SdS}}
= \tilde \Phi -(1/2) (\ddot{a}/a) R^2$ satisfies the equation
%%%%%%
\be{5.19} \triangle_{{\bf R}}\Phi_{\mathrm{SdS}}= \triangle_{{\bf R}}\tilde \Phi -3 \frac{\ddot{a}}{a}= 4\pi G_N\left(\rho_{\mathrm{ph}}-\frac{\Lambda
c^2}{4\pi G_N}\right)\, . \ee
%%%%%%
If we are not making any additional assumptions about the distribution of gravitating masses and the boundary conditions for the gravitational potential far
from them, then in the case $\rho_{\mathrm{ph}}=\sum_i m_i\delta({\bf R}-{\bf R}_i)$, the solution of this equation can be formally written as
%%%%%%%%
\be{5.20}
\Phi_{\mathrm{SdS}}=-G_N\sum\limits_{i=1}^N\frac{m_i}{|{\bf R}-{\bf R}_i|}-\frac{\Lambda c^2R^2}{6}\, ,
\ee
%%%%%%%%
where we dropped an arbitrary function of time. The drawback of such a potential is the lack of translational symmetry. In other words, the result depends on
the choice of the origin. For the finite number of gravitating masses, we can restore this symmetry by rewriting the solution as follows:
%%%%%%%
\be{5.21}
\Phi_{\mathrm{SdS}}=-G_N\sum\limits_{i=1}^N\frac{m_i}{|{\bf R}-{\bf R}_i|}-\frac{\Lambda c^2}{6}
\sum\limits_{i=1}^N\frac{m_i({\bf R}-{\bf R}_i)^2}{\sum\limits_{i=1}^Nm_i}\, .
\ee
%%%%%%%
The prefactors $m_i/\sum\limits_{i=1}^Nm_i$ are chosen in such form to preserve the Newtonian third law. However, in the case of infinite number of masses (as
it takes place in spatially flat Universe), the second term vanishes. Moreover, the sum of an infinite number of Newtonian potentials diverges (the
Neumann-Seeliger paradox \cite{Norton}).

It is worth to make a couple of additional comments on gravitational potentials in the model $\mathcal{K}=0$. It can be easily seen that the potential
$\Phi_{\mathrm{SdS}}$ in \rf{5.20} diverges at $R\to \infty$ (where the gravitating mass relating to this potential is absent). However, any nonrelativistic
gravitational potential should be less than $c^2$. Clearly, because of extreme smallness of $\Lambda$, the term $\Lambda R^2$ becomes greater than 1 at very
large distances, but we cannot apply this potential to the  whole infinite Universe. In contrast to $\Phi_{\mathrm{SdS}}$, the potential $\varphi$ (or $\tilde
\Phi$) in \rf{5.11}, \rf{5.12} is finite everywhere outside the gravitating mass. However, for the considered artificial distribution of the inhomogeneities,
the average value of the potential of all inhomogeneities over the whole Universe is not equal to zero: $\overline \varphi \neq 0$ \cite{EZJCAP}. It
contradicts to the natural assumption that in models, where the average density fluctuation $\overline {\delta\rho}=\overline {\rho -\overline\rho}=0$, the
average gravitational potential $\overline\varphi$ must also vanish (because $\overline {\delta\rho}$ is a source of $\overline\varphi$). Thus, the model with
such distribution of gravitating masses is not satisfactory. Above, we mentioned that the periodic distribution of inhomogeneities \cite{periodic} also has
problematic aspects.  Therefore, the $\mathcal{K}=0$ model require a rather specific distribution of inhomogeneities to overcome all these
problems\footnote{\label{5}There is a possibility to avoid these problems in the $\mathcal{K}=0$ model if we include an exotic matter, e.g., in the form of a
frustrated network of cosmic strings \cite{BUZ1}.}.

As we have shown in \cite{EZJCAP}, the hyperbolic $\mathcal{K}=-1$ model is free of all these drawbacks. First, the gravitational potential $\varphi$ preserves
the translational symmetry and satisfies the principle of superposition (see Eq. (4.7) in \cite{EZJCAP}). It is finite everywhere outside gravitating masses
for an arbitrary distribution of inhomogeneities. Moreover, it converges rapidly at large distances from the gravitating mass due to the corresponding
Yukawa-type behavior (that allows to resolve the Neumann-Seeliger paradox) and its average value is equal to zero. Second, the Lagrange equation for the
inhomogeneities inside the cell of uniformity reads \cite{EZK2}
%%%%%%
\be{5.22}
\frac{d{\bf V}_i}{dt} = \frac{\ddot{a}}{a}{\bf R}_i -G_N \sum_{j\neq i}\frac{m_j({\bf R}_i-{\bf R}_j)}{|{\bf R}_i-{\bf R}_j|^3}\, .
\ee
%%%%%%%
This equation shows that at large distances from inhomogeneities we have the correct asymptote \rf{5.18}. Additionally, the peculiar acceleration here is the
sum of the Newtonian expressions as it is claimed in \cite{Peebles} (see Eq. (8.5) in \cite{Peebles}). Exactly these equations are used for the N-body
simulation \cite{gadget2} (with the appropriate smoothing of delta-shaped gravitating sources).

%%%%%%%%%%%%%%%%%%%%%%%%%%%%%%%%%%%%%%%%%%%%%%%%%%%%%%%%%%%%%%%%%

\section{\label{sec:6}Conclusion}

In this paper, we have considered the mechanical approach to the observable Universe. This approach was proposed in our previous paper \cite{EZJCAP} and here
we continued its further justification. The reason for this approach is that, according to observations, at scales smaller than 100-300 Mpc the Universe is
highly inhomogeneous and structured. We observe isolated galaxies, which form clusters and superclusters. Obviously, these isolated inhomogeneities can not be
represented in the form of a fluid. Therefore, hydrodynamics is not appropriate to describe their behavior on the considered scales. We need to create a
mechanical approach where dynamical behavior is defined, on the one hand, by gravitational potentials of inhomogeneities and, on the other hand, by the
cosmological expansion of the Universe. On much bigger scales the Universe becomes on average homogeneous and isotropic with matter in the form of a perfect
fluid. In the present paper, we, first, estimated scales at which the transition from the mechanical approach to the hydrodynamical one occurs. To perform this
estimate, we considered the idealized Universe filled with the typical galaxies of the mass and size of the Milky Way or Andromeda. Then, using the standard
methods of statistical physics, we found that the cell of uniformity size is approximately equal to 190 Mpc, which is rather close to observations. This
estimate depends on the parameters of the chosen typical galaxy, i.e. on its size and mass. If we take, e.g., a typical galaxy in ten times smaller (in the
radius) and in two orders of magnitude lighter than MW and M31, we get 76 Mpc for the cell of uniformity size, which is also rather big.

To describe the evolution of the inhomogeneities inside the cell of uniformity, we must apply the mechanical approach. This approach within the $\Lambda$CDM
model was introduced briefly in section 3. We stressed here that we can apply this method in the case of nonrelativistic peculiar velocities $v_{\mathrm{ph}}$.
If we demand that $|v_{\mathrm{ph}}/c|< 10^{-2}$, then, according to our calculations, this condition works for the redshifts $z\lesssim 10$ which correspond
approximately to 13 billion years from the present moment, i.e. almost the entire age of the Universe!

In the zero order approximation when we drop gravitational perturbations caused by the inhomogeneities and do not take into account their peculiar velocity,
the Universe is described by the background Friedmann model. Inhomogeneities result in perturbation of this background. In section 4, we have considered scalar
perturbations. It is important to stress that we considered the gravitational potential and the peculiar velocity of inhomogeneities as two independent small
parameters. A similar approach to the gravitational field of an arbitrary number of astrophysical objects in the weak field limit was considered in
\cite{Landau}. It was demonstrated here that the peculiar velocities in the first order approximation do not affect the gravitational potentials. In turn, the
gravitational potentials define the dynamics of test bodies. Therefore, we consider the theory of scalar perturbations in the first order with respect to the
gravitational potentials $\Phi$ and in the zero order concerning the peculiar velocities $\tilde v$. In this case all terms/perturbations in Einstein equations
are of the order of $O(1/c^2)$. We demonstrated that account of the peculiar velocities leads to additional terms which are product of two small parameters
$\Phi$ and $\tilde v$ and these terms are of the order of $O(1/c^4)$ or $O(1/c^3)$. Therefore, within the accuracy of our approach we can drop them. This led
us to the master equation \rf{4.11} for the gravitational potential of inhomogeneities. We have also demonstrated that radiation can be naturally incorporated
into our scheme. This emphasizes the viability of our approach.

Having the gravitational potential at hand, we can construct the Lagrange function for a test body and investigate its dynamics taking into account both
gravitational interaction between inhomogeneities and cosmological expansion of the Universe. Such Lagrange function takes the form of Eq. \rf{5.2}. It is
worth noting that both the master equation \rf{4.11} for the gravitational potential (in the case of the flat Universe $\mathcal{K}=0$) and the Lagrange
function \rf{5.2} exactly coincide with the corresponding formulas in \cite{Peebles,gadget2}. It is important to note also that in our case these equations
were obtained from the first principles. This enables us to generalize our analysis on the various alternative cosmological models and check their
compatibility with observations (see, e.g., \cite{BUZ1}). This is an important advantage of our approach.

At first glance, the mechanical approach should not describe the growth of structure. However, we have shown that in the Newtonian approximation, starting from
our equations, we get the standard formulas that describe the growth of the density contrast.

Then, we discussed some problematic aspects of the flat model ($\mathcal{K}=0$). The main problem here is that the inhomogeneities should undergo a rather
specific spatial distribution in the Universe. Otherwise, the gravitational potential produced by all inhomogeneities in the infinite Universe can diverge
outside the gravitating bodies. The Schwarzschild-de Sitter solution is one of such examples. If we do not suppose any specific boundary conditions outside
gravitating masses and apply this solution for the entire infinite Universe, then the gravitational potential produced by all inhomogeneities will diverge at
any points outside these masses. Another important drawback of this solution is that the Universe at large scales is the de Sitter one but not the Friedmann
Universe. Therefore, it does not take into account the matter in the Universe which contributes 31 \% into the total balance. This is the accuracy of this
approach. Thus, the Schwarzschild-de Sitter solution should not be used to describe the motion of test bodies in the cosmological background. In the case of
the flat space $\mathcal{K}=0$, we also presented two particular examples with unphysical properties. For the model with the periodic distribution of the
gravitating masses, the gravitational potential diverges at points where masses are absent. Next, we considered the model where the potential is finite
everywhere in the space but its average value is nonzero. It contradicts to the natural condition that the average value of the energy density fluctuations
must be zero.

We have shown that, within the $\Lambda$CDM model, the hyperbolic Universe ($\mathcal{K}=-1$) is free of the drawbacks inherent in the spatially
flat Universe \cite{EZJCAP}. However, in alternative (with respect to the $\Lambda$CDM one) models, physically reasonable solutions can take place for any sign
of the spatial curvature $\mathcal{K}$ \cite{BUZ1}.

Hence, in this paper we have provided additional evidence in favor of the mechanical approach to the Universe inside the cell of uniformity. We have also
demonstrated the application of this method for analyzing different cosmological models.

%%%%%%%%%%%%%%%%%%%%%%%%%%%%%%%%%%%%%%%%%%%%%%%%%%%%%%%%%%%%%%%%%%%%%%%%%%%

%\vspace{0.5cm}

\acknowledgments

The work of M. Eingorn was supported by NSF CREST award HRD-1345219 and NASA grant NNX09AV07A.

We thank Yu. Shtanov, B. Novosyadlyj, V. Kulinskii and F. Labini for stimulating discussions and critical comments.
%We want to
%thank the referee for his/her comments which have considerably improved the motivation of our investigations and the presentation of the %results.

%%%%%%%%%%%%%%%%%%%%%%%%%%%%%%%%%%%%%%%%%%%%%%%%%%%%%%%%%%%%%%%%%%%%%%%%%%%%%%%%%%%%%%%%%%%%%%%%%%%%%%%%%%%%%%%%%
%%%%%%%%%%%%%%%%%%%%%%%%%%%%%%%%%%%%%%%%%%%%%%%%%%%%%%%%%%%%%%%%%%%%%%%%%%%%%%%%

\end{document}